\documentclass[preprint,showpacs,preprintnumbers,amsmath,amssymb,eqsecnum]{revtex4}
\usepackage{graphicx}
\usepackage{dcolumn}
\usepackage{bm}
\newcommand{\mathsym}[1]{{}}

\newcommand{\ba}{\mbox{\boldmath $a$}}
\newcommand{\bb}{\mbox{\boldmath $b$}}
\newcommand{\bc}{\mbox{\boldmath $c$}}

\begin{document}
 
\title{Domain wall solution of the Skyrme model\\}
 
\author{Chang-Guang Shi}
 \email{shichangguang@shiep.edu.cn}
\affiliation{Department of Mathematic and Physics, Shanghai University of Electric power, Pingliang Road 2103,
 Shanghai 200090, {\bf China}\\}
\author{Minoru Hirayama}
 \email{hirayama@sci.u-toyama.ac.jp}
\affiliation{Department of Physics,  University of Toyama, Gofuku 3190,
 Toyama 930-8555, {\bf Japan}\\}
                                                                                
\begin{abstract}
A class of domain-wall-like solutions of the Skyrme model is obtained analytically. They are described by the tangent hyperbolic function,  which is a special limit  of  the Weierstrass $\wp$ function.  The behavior of one of the two terms in the static energy density is  like that of  a domain wall. The other term in the static energy density does not vanish but becomes constant at the points far apart from the wall.
\end{abstract}

\maketitle

\section{Introduction}\label{sec:1}
                                                                                
Skyrme Model  \cite{Skyrme} is an effective
field theory describing hadrons \cite{Witten}, \cite{Jackson}. It is defined by
the Lagrangian density
\\
\begin{align}
{\cal L}_S=&-4c_2
 \mbox{tr}\left[(g^{\dagger}\partial_{\mu}g)(g^{\dagger}\partial^{\mu}g)\right
]+\frac{c_4}{2}\mbox{tr}\left([g^{\dagger}\partial_{\mu}g,g^{\dagger}\partial_{\nu}g][g^{
\dagger}\partial^{\mu}g,g^{\dagger}\partial^{\nu}g]\right),
\label{eqn:Lagrangian}
\end{align}
where $g(x)$ is an element of $SU$(2) and $c_2$ and $c_4$ are coupling
constants. If we define
$A_{\mu}^{\alpha}(x)$ and $H_{\mu\nu}^{\alpha}(x)$ by
\begin{align}
&A_{\mu}^{\alpha}=\frac{1}{2i}\mbox{tr}\left(\tau^{\alpha}g^{\dagger}\partial_{\mu}g\right),
\label{eqn:Amug1} \\
&H_{\mu\nu}^{\alpha}=\partial_{\mu}A_{\nu}^{\alpha}-\partial_{\nu}A_{\mu}^{\alpha},
\end{align}
${\cal L}_S$ is expressed as
\begin{align}
{\cal
 L}_S=8c_2A_{\mu}^{\alpha}A^{\alpha,\mu}-c_4H_{\mu\nu}^{\alpha}H^{\alpha,\mu\nu},
\label{eqn:LagrangianA}
\end{align}
where $\tau_{\alpha} (\alpha=1, 2, 3)$ are Pauli matrices.
By definition, $A_{\mu}^{\alpha}(x)$ satisfies the condition
\begin{align}
\partial_{\mu}A_{\nu}^{\alpha}-\partial_{\nu}A_{\mu}^{\alpha}=2\varepsilon^{\alpha\beta\gamma}A
_{\mu}^{\beta}A_{\nu}^{\gamma}.
\label{eqn:conditiona}
\end{align}
The field
equation is given as the
conservation law
\begin{align}
\partial_{\mu}J^{\alpha, \mu}=0,
\label{eqn:conservationlow}
\end{align}
where $J^{\alpha, \mu}$ $(\alpha=1, 2, 3,\hspace{2mm}\mu=0, 1, 2, 3)$
are defined by
\begin{align}
J^{\alpha, \mu}=2c_2A^{\alpha,
 \mu}+c_4\varepsilon^{\alpha\beta\gamma}H^{\beta, \mu\nu}A_{\nu}^
{\gamma}.
\label{eqn:Jmu}
\end{align}
The field $J^{\alpha, \mu}(x)$ is proportional to the isospin current of
the model.
 Another important current of the Skyrme model is the baryon number current $N
^\lambda(x)$ \cite{Skyrme} defined by
\begin{align}
N^{\lambda}=\frac{1}{12\pi^2}\varepsilon^{\lambda\mu\nu\rho}\varepsilon^{\alpha\beta\gamma}A_{
\mu}^{\alpha}A_{\nu}^{\beta}A_{\rho}^{\gamma}.
\label{eqn:BaryonN}
\end{align}
The conservation law $\partial_{\lambda}N^{\lambda}=0$ follows solely
from the definition of $N
^{\lambda}(x)$ irrespective of the field equation for
$A_{\mu}^{\alpha}(x)$.
It was  shown that solitons of the Skyrme model possess polyhedral structures \cite{Battye}, \cite{Sutcliffe}.
As for the analytic solutions for these models, only a few simple
examples \cite{Skyrme,Hirayama,HSY} are known. Skyrme \cite{Skyrme} found that the configuration
\begin{align}
g(x)=h(k\cdot x)
\label{eqn:kf}
\end{align}
with $k\cdot x=k_{\mu}x^{\mu}$ leading to $A_{\mu}^{\alpha}(x)=k_{\mu}f^{\alpha}(k\cdot x)$ satisfies the field equation if $k_{\mu}$ is light-like: $k^2=0$.  For these solutions, the field $H^{\alpha}_{\mu\nu}(x)$ vanishes and  $g(x)$ is independent of the coupling constants $c_2$ and $c_4$. In a recent paper \cite{HSY}, Yamashita and the present authors  obtained a solution of the form
\begin{align}
g(x)=h(k\cdot x,\hspace{2mm} l\cdot x, \hspace{2mm}m\cdot x),
\label{eqn:klm}
\end{align}
with $k$, $l$, $m$ being three momenta satisfying $k^2 = l^2 = m^2 = 0$.
In that case, the field $A^{\alpha}_{\mu}(x)$ is expressed as
\begin{equation}
A^{\alpha}_{\mu}(x) = \frac{k_{\mu}}{\kappa^1}a^{\alpha}(\omega,\omega ')+\frac{l_{\mu}}{\kappa^2}b^{\alpha}(\omega,\omega ') +\frac{m_{\mu}}{\kappa^3}c^{\alpha}(\omega,\omega '),\\
\label{eqn:Aam}
\end{equation}
where the variables  $\omega$ and $\omega'$ are linear in $x^\mu$ and $\kappa^i$ is defined as
\begin{align}
\kappa^i&=\sqrt{\frac{c_4}{c_2}\frac{(k^i\cdot k^j)(k^i\cdot
 k^k)}{(k^j\cdot k^k)}},
\label{eqn:par}
\end{align}
with the triplet $(i,j,k)$  being (1,2,3) or (2,3,1) or (3,1,2) and
$k^1=k$, $k^2=l$, $k^3=m$.
 In this case, we can see that the solutions dependent on  $c_2, c_4$ nontrivially, and that the field
$H^{\alpha}_{\mu\nu}(x)$ and the baryon number density are  nonvanishing.
Under the Ansatz used in ref. \cite{HSY}, the quantities $|{\bm{a}}|$, $|{\bm{b}}|$ and $|{\bm{c}}|$ are constants and $\bm{a\cdot b}$, $\bm{b\cdot c}$ and $\bm{c\cdot a}$ are described with the help of the function
\begin{align}
K(\omega)&=\wp(\omega+\omega_3) \nonumber \\
&=e_3+\frac{(e_3-e_1)(e_3-e_2)}{\wp(\omega)-e_3} \label{eqn:Komega}\\
&=e_3+(e_2-e_3)\mbox{sn}^2\left(\sqrt{e_1-e_3}\hspace{1mm}\omega,
 \,\sqrt{\frac{e_2-e_3}{e_1-e_3}}
\right), \nonumber
\end{align}
where $\wp(z)$ is the Weierstrass $\wp$ function satisfying the
differential equation
\begin{align}
\left[\wp^{\prime}(z)\right]^2=4\left[\wp(z)-e_1\right]\left[\wp(z)-e_2\right]\left[\wp(z)-e_3\right].
\label{eqn:wpfunc}
\end{align}
Here the constants $e_1, e_2$ and $e_3$ are complicated functions of $|{\bm{a}}|$, $|{\bm{b}}|$ and $|{\bm{c}}|$. They are real and are assumed to satisfy $e_1>e_2>e_3$.
sn$(u, k)$ is the Jacobi
elliptic function of $u$ with the modulus $k$, $2\omega_3$ is the second
fundamental period of $\wp(
z)$, and $\omega=L\cdot x$ is a linear combination of $k\cdot x,
\hspace{2mm}l\cdot x$ and $m\cdot
x$. We note that $L^2$ is equal to $c_2/c_4$ multiplied by a constant
 independent of the momenta
$k$, $l$ and $m$. The function $K(\omega)$  oscillates even for large $|{\omega}|$.Noting that the static energy density and the baryon number
density are described by $K(\omega)$ and its derivative,respectively, we see that they also oscillate at the spatial infinity.
                                                                                
In this paper, we consider the limit that the function $\rm{sn}(u,k)$ appearing
in Eq. (\ref{eqn:Komega}) tends to $\rm{Tanh}(u)$. Then the simple behavior of $\rm{Tanh}(u)$ suggests  that we might obtain a domain wall configuration in the
limit. This limit is realized if the parameter  $e_1$ tends to $e_2$. Because of the complicated dependency of  $(e_1, e_2, e_3)$ on $|{\bm{a}}|$, $|{\bm{b}}|$
and $|{\bm{c}}|$, it is necessary to show that there indeed exists a set $( |{\bm{a}}|, |{\bm{b}}|, |{\bm{c}}|)$ realizing $e_1 \rightarrow e_2$ and $K(\omega)
\rightarrow e_3+(e_2-e_3)\left[\rm{Tanh}(\sqrt{e_2-e_3} \omega)\right]^2$.
                                                                                
As will be shown in later sections, this is the case. As is seen from Eq. (\ref{eqn:Lagrangian}), the static energy density of the Skyrme model consists of two
terms: the one quadratic and the other quartic in field variables. It turns out
that the behavior of the quartic term is like that of a domain
wall. It approaches to zero at points far apart from the wall. The behavior of the quadratic term is  also like that of a domain wall. It approaches to a constant at points far apart from the wall. It should be noted that the last constant
is non-vanishing  The baryon number density concentrates near the wall. The total baryon number, however, vanishes.
                                                                                
This paper is organized as follows. In \ref{sec:2},
 we briefly introduce the solutions of the Skyrme model described by the $\wp$ function.  In \ref{sec:3}, we explain how to explore domain wall solutions. the detail of the  analysis is given. \ref{sec:4} is devoted to a summary and discussions.
\section{Solutions described by $\wp$ function}\label{sec:2}
\subsection{Field equation}
To be self-contained, we briefly review the method of ref. \cite{HSY} in this section.
Introducing $\xi^i$ by
\begin{equation}
\xi^i=\frac{k^i\cdot x}{\kappa^i}\hspace{5mm}(i=1,2,3)
\label{eqn:xi}
\end{equation}
and write $\xi=\xi^1$, $\eta=\xi^2$ and $\zeta=\xi^3$, we find that the integrability condition
(\ref{eqn:conditiona}) yields
\begin{align}\begin{array}{ll}
\displaystyle
{\frac{\partial{\bm a}}{\partial\eta}-\frac{\partial{\bm b}}{\partial\xi}}&=2({\bm b}\times{\bm a}),\\
\displaystyle
{\frac{\partial{\bm a}}{\partial\zeta}-\frac{\partial{\bm c}}{\partial\xi}}&=2({\bm c}\times{\bm a}),
\\
\displaystyle
{\frac{\partial{\bm b}}{\partial\zeta}-\frac{\partial{\bm c}}{\partial\eta}}&=2({\bm c}\times{\bm b}).
\end{array}\label{eqn:condition}
\end{align}
Here, $\bm a, \bm b$ and $\bm c$ are three-dimensional vectors introduced in Eq.(1.11). If we use $\xi, \eta$, and $\zeta$, the field equation  (\ref{eqn:conservationlow}) becomes
\begin{align}
&\left(\frac{\partial}{\partial\eta}+\frac{\partial}{\partial\zeta}\right)\left({\bm a}+{\bm D}\right)+
\left(\frac{\partial}{\partial\xi}+\frac{\partial}{\partial\zeta}\right)\left
({\bm b}+{\bm E}\right)+\left(
\frac{\partial}{\partial\xi}+\frac{\partial}{\partial\eta}\right)\left({\bm c
}+
{\bm F}
\right)=0,
\label{eqn:QMotion1}
\end{align}
where ${\bm D}$, ${\bm E}$, ${\bm F}$ are given by
\begin{align}
{\bm D}&={\bm a}\times({\bm b}\times{\bm a})+{\bm a}\times({\bm c}\times{\bm a})+{\bm c}\times({\bm b}\times{\bm a})+{\bm b}\times({\bm c}\times{\bm a}),\nonumber\\
{\bm E}&={\bm b}\times({\bm a}\times{\bm b})+{\bm b}\times({\bm c}\times{\bm b})+{\bm c}\times({\bm a}\times{\bm b})+{\bm a}\times({\bm c}\times
{\bm b}), \label{eqn:DEF}\\
{\bm F}&={\bm c}\times({\bm a}\times{\bm c})+{\bm c}\times({\bm b}\times{\bm c})+{\bm a}\times({\bm b}\times{\bm c})+{\bm b}\times({\bm a}\times
{\bm c}).\nonumber
\end{align}
To solve this highly  nonlinear field equation, we must introduce some Ans\"atze.
\subsection{Ans\"atze}
The Ans\"atze adopted in ref. \cite{HSY} is summarized as
 \begin{align}
\left(
\begin{array}{ccc}
{\displaystyle\frac{\partial  \mbox{\boldmath{$a$}}}{\partial \xi}} & {\displaystyle\frac{\partial  \mbox{\boldmath{$a$}}}{\partial \eta}} & {\displaystyle\frac{\partial  \mbox{\boldmath{$a$}}}{\partial \zeta}}\\
{\displaystyle\frac{\partial  \mbox{\boldmath{$b$}}}{\partial \xi}}  & {\displaystyle\frac{\partial  \mbox{\boldmath{$b$}}}{\partial \eta}} &{\displaystyle \frac{\partial  \mbox{\boldmath{$b$}}}{\partial
\zeta}}\\
{\displaystyle\frac{\partial  \mbox{\boldmath{$c$}}}{\partial \xi}} & {\displaystyle\frac{\partial  \mbox{\boldmath{$c$}}}{\partial \eta}} &{\displaystyle \frac{\partial  \mbox{\boldmath{$c$}}}{\partial \zeta}}
\end{array}
\right) = & \left(
\begin{array}{ccc}
\lambda & \gamma& 0\\
\gamma+2 & -\rho &0\\
0&0&0
\end{array}
\right)
( \mbox{\boldmath{$a$}}\times  \mbox{\boldmath{$b$}})+
\left(
\begin{array}{ccc}
0 & 0& 0\\
0 & \sigma & \alpha\\
0& \alpha +2& -\nu
\end{array}
\right)
( \mbox{\boldmath{$b$}}\times  \mbox{\boldmath{$c$}})\nonumber\\
& +
\left(
\begin{array}{ccc}
-\kappa & 0& \beta+2\\
0 & 0 & 0\\
\beta& 0& \mu
\end{array}
\right)
( \mbox{\boldmath{$c$}}\times  \mbox{\boldmath{$a$}}),
\end{align}
where $\alpha,\beta, \gamma, \kappa, \lambda, \mu, \nu, \rho$ and $\sigma$ are
constants.
It was found in ref. \cite{HSY}  that the Ans\"atze is consistent with the integrability condition (\ref{eqn:condition}) and the field equation (\ref{eqn:QMotion1})
 if the parameters satisfy
\begin{align}
\rho=\frac{\mu}{\nu} \frac{\alpha (\alpha+2)}{\alpha+\beta+2}, \quad
&\lambda=-\frac{\nu}{\mu} \frac{\beta (\beta+2)}{\alpha+\beta+2}, \quad
\gamma=- \frac{(\alpha+2) (\beta+2)}{\alpha+\beta+2}, \nonumber\\
& \kappa=-\frac{\beta (\beta+2)}{\mu} , \quad \sigma=-\frac{\alpha (\alpha+2)}{\nu}.
\end{align}
Therefore we can regard the four constants $\alpha,\beta,\mu$ and $\nu$  as independent.
\subsection{Solution of the Ans\"atze}
From the above Ans\"atze, we see  that $\bm{a}^2$, $\bm{b}^2$ and $\bm{c}^2$ are constants.  We also obtain
\begin{align}
&\ba\cdot\bb=(\alpha+\beta+2)J(\omega)+d_1, \label{eqn:ab}\\
&\bb\cdot\bc=\mu J(\omega)+d_2 \label{eqn:bc}, \\
&\bc\cdot\ba=-\nu J(\omega)+d_3 \label{eqn:ca}, \\
&\ba\cdot\left(\bb\times\bc\right)=-\alpha(\alpha+2)\frac{\mu}{\nu}
\frac{dJ(\omega)}{d\omega},
\label{eqn:p} \\
&\omega=\rho\kappa\xi-\rho\sigma\eta+\mu\sigma\zeta,  \label{eqn:omega}
\end{align}
where $J(\omega)$ is a function of $\omega$ and $d_1, d_2$ and $d_3$ are arbitrary constants.
With the help of the identity
\begin{align}
\left[  \mbox{\boldmath{$a$}} \cdot ( \mbox{\boldmath{$b$}} \times  \mbox{\boldmath{$c$}})\right]^2 =
 \left|
\begin{array}{ccc}
\mbox{\boldmath{$a$}}^2 & ( \mbox{\boldmath{$a$}} \cdot  \mbox{\boldmath{$b$}}) & ( \mbox{\boldmath{$a$}} \cdot  \mbox{\boldmath{$c$}})\\
( \mbox{\boldmath{$b$}} \cdot  \mbox{\boldmath{$a$}}) &  \mbox{\boldmath{$b$}}^2 & ( \mbox{\boldmath{$b$}} \cdot  \mbox{\boldmath{$c$}})\\
( \mbox{\boldmath{$c$}} \cdot  \mbox{\boldmath{$a$}}) & ( \mbox{\boldmath{$c$}}
\cdot  \mbox{\boldmath{$b$}}) &  \mbox{\boldmath{$c$}}^2
\end{array}
\right|,
\end{align}
it is concluded that $J(\omega)$ is related to the function $K(\omega)$ in Eq. (\ref{eqn:Komega}) by
\begin{align}
&K(\omega)=\frac{1}{z_1}J(\omega)-\frac{z_2}{z_1}, \\
&z_1=-\frac{2\alpha^2(\alpha+2)^2\mu}{(\alpha+\beta+2)\nu^3}, \\
&z_2=\frac{-\mu^2\ba^2-\nu^2\bb^2-(\alpha+\beta+2)^2\bc^2-2\mu\nu
 d_1+2(\alpha+\beta+2)
(\mu d_3-\nu
 d_2)}{6(\alpha+\beta+2)\mu\nu}.
\label{eqn:Jz1z2}
\end{align}
As is seen from (\ref{eqn:Komega}) and (\ref{eqn:wpfunc}), $K(\omega)$ satisfies                                                                                
\begin{align}
\left[\frac{d K(\omega)}{d \omega} \right]^2=& 4 \left[K(\omega) \right]^3 -g_2
K(\omega) - g_3\nonumber\\
=& 4 \left[ K(\omega) -e_1\right]  \left[ K(\omega) -e_2\right]  \left[ K(\omega) -e_3\right],
\end{align}
where $g_2, g_3, e_1, e_2$, and $e_3$ are real constants satisfying  $e_1 \geq e_2 \geq e_3$. They are related by
\begin{align}
e_1+ e_2+ e_3 = 0,\quad  e_2 e_3 + e_3 e_1 + e_1 e_2 = - \frac{1}{4} g_2,\quad e_1 e_2 e_3 =  \frac{1}{4} g_3.
\end{align}
If we set $K(\omega) = \wp (\omega + \rm{const}) = \wp(\omega+\omega_3)$  with
$\omega_3 = \frac{i}{2} \int_{-\infty}^{e_3} \frac{d u}{\sqrt{(e_1 - u) ( e_2 -
u) ( e_3 - u)}}$,
we find that $K(\omega)$  is regular for real values of $\omega$ and expressed
by the Jacobi elliptic function by (\ref{eqn:Komega}).
                                                                                
\subsection{Solution of the field equation}
The field equation is now reduced to six algebraic equations.
Three of them are solved by fixing $ d_1,d_2, d_3$ as
\begin{align}
\left(
\begin{array}{c}
d_1\\
d_2\\
d_3
\end{array}
\right)
&= \frac{1}{2ABC}
\left(
\begin{array}{c}
-C(As_0 +Bt_0 - Cu_0)\\
A(As_0-Bt_0-Cu_0)\\
-B(As_0 - Bt_0+ Cu_0)
\end{array}
\right),\\
& A=(\alpha+\beta+\gamma)-3\alpha+(\nu-\mu)+(\rho-\sigma),\\
& B=(\alpha+\beta+\gamma)-3\beta+(\kappa-\lambda)+(\nu-\mu),\\
& C=(\alpha+\beta+\gamma)-3\gamma+(\rho-\sigma)+(\kappa-\lambda),\\
& s_0=(\beta+\gamma+2+\rho-\mu) \mbox{\boldmath{$a$}}^2-\nu \mbox{\boldmath{$b$}}^2+\sigma \mbox{\boldmath{$c$}}^2+2(\alpha+1),\\
& t_0=\mu \mbox{\boldmath{$a$}}^2+(\alpha+\gamma+2+\nu-\lambda) \mbox{\boldmath{$b$}}^2-\kappa \mbox{\boldmath{$c$}}^2+2(\beta+1),\\
& u_0=-\rho \mbox{\boldmath{$a$}}^2+\lambda \mbox{\boldmath{$b$}}^2+(\alpha+\beta+2+\kappa-\sigma) \mbox{\boldmath{$c$}}^2+2(\gamma+1).
\end{align}
The other three equations are reduced to the restrictions on the parameters $(\alpha,\beta,\mu,\nu)$.
They can be written in the form $p_i \mu^2+2h_i\mu\nu+q_i\nu^2+2 g_i+2f_i\nu+r_i=0~ (i=1,2,3)$,where $p_i,q_i,,r_i,f_i,g_i$ and $h_i$ are polynomials of $\alpha$ and $\beta$. They were solved in ref. \cite{HSY}.
Two examples of the allowed sets are given by
 \begin{align}
(\alpha=\beta,\mu,\nu)=\left(-2-\frac{1}{\sqrt{2}}+\frac{\sqrt{5+4\sqrt{2}}}{2},1,-1 \right)
\end{align}
and
 \begin{align}
(\alpha=\beta,\mu,\nu)=\left(-2-\frac{1}{\sqrt{2}}-\frac{\sqrt{5+4\sqrt{2}}}{2},1,-1 \right).
\end{align}
\section{Exploring domain wall solutions}\label{sec:3}
\subsection{Conditions to realize a domain wall}
Recalling that $K(\omega)$ contains the factor $(e_2 - e_3) \left[ \rm{sn}\left( \sqrt{e_1 -e_3}\hspace{1mm}  \omega, \sqrt{\frac{e_2 - e_3}{e_1 - e_3}} \right) \right]^2$, we here consider some limiting cases of this factor.
From the definition
\begin{align}
& u = \int^{{\rm{sn}} (u,k)}_0 \frac{dz}{\sqrt{(1 - z^2) (1 - k z^2)}}
\end{align}
of the Jacobi elliptic function, we easily find
\begin{align}
& {\rm{sn}} (u,0) = \sin u,\quad {\rm{sn}} (u,1) = {\tanh} u.
\end{align}
The  $\rm{sn} \rightarrow \sin$ limit in our case is realized by $e_2 \rightarrow e_3$ and leads to trivial $K(\omega)$ : $K(\omega) = e_3$. On the other hand,
the  $\rm{sn} \rightarrow \tanh$ limit is realized by
\begin{align}
&e_1\rightarrow e_2,\quad e_3\rightarrow -2e_2
\end{align}
and leads to nontrivial $K(\omega)$:
\begin{align}
&K(\omega) = -2 e_2 + 3e_2 [ \tanh\chi]^2,\quad \chi =\sqrt{3 e_2} \hspace{1mm}
\omega.
\end{align}
From the property of the function $[\tanh\chi]^2$, we expect  that a domain wall-like structure might appear in this limit. Since $e_1$, $e_2$ and $e_3$  depend on the parameters $\alpha$, $\beta$, $\mu$, $\nu$ and on the constants $\bm{a^2}$, $\bm{b^2}$, $\bm{c^2}$ in a complicated manner, we have to check weather  this limit can be realized or not.
                                                                                
\subsection{A special case}
We consider the case
\begin{equation}
\mu=1,\quad  \nu=-1,\quad  \alpha=\beta=-2-\frac{1}{\sqrt{2}}+\frac{\sqrt{5+4\sqrt{2}}}{2} = - 0.398768\cdots
\end{equation}
and assume that  $\bm{k}$, $\bm{l}$, $\bm{m}$ are orthogonal to each other.
The conditions to be considered are summarized as

\begin{align}
&\hspace{1cm}(1)\hspace{2mm} e_1=e_2, ~ e_3=-2e_2\nonumber\\
&\hspace{2.7cm}\rightarrow  g_2= 12(e_2)^2,\quad g_3=- 8(e_2)^3.\\
&\hspace{2.7cm}\rightarrow g_2^3 = 27 g_3^2,\nonumber\\
&\hspace{1cm} (2)\hspace{2mm} e_2 > 0\rightarrow g_3<0.
\\
&\hspace{1cm} (3)\hspace{2mm} {\bm{a}}^2  {\bm{b}}^2 \geq ( {\bm{a}}\cdot {\bm{b}})^2,\quad {\bm{b}}^2  {\bm{c}}^2 \geq ( {\bm{b}} \cdot {\bm{c}})^2, \quad
{\bm{c}}^2  {\bm{a}}^2 \geq ( {\bm{c}} \cdot {\bm{a}})^2  \hspace{1mm}\text{for}\hspace{1mm} \text{any} \hspace{1mm}\chi. \label{eqn:cond3}
\end{align}
We note that ($ \mbox{\boldmath{$a \cdot b$}})$ , ($ \mbox{\boldmath{$b \cdot c$}})$,  ($ \mbox{\boldmath{$c \cdot a$}})$ are linear in $ \mbox{\boldmath{$a^2,
b^2, c^2$}}$ and contain $[ \tanh\chi]^2$, $g_2$ is a  polynomial of $ \mbox{\boldmath{$a^2, b^2, c^2$}}$ of second order,  $g_3$ is  a polynomial of $ \mbox{\boldmath{$a^2, b^2, c^2$}}$ of third order, $g_2^3 - 27 g_3^2$ is a polynomial of \mbox{\boldmath{$a^2, b^2, c^2$}} of sixth order.
\subsection{Static energy density}
The energy momentum tensor is given by
\begin{equation}
T_{\mu\nu}=\frac{\partial \mathcal{L}_s}{\partial A^{\alpha,\mu}}A^\alpha_{\nu}-\eta_{\mu\nu}\mathcal{L}_s.
\label{eqn:energymt}
\end{equation}
Setting
\begin{align}
&k=(k_0, k_0,0,0),\quad  l=(l_0,0,l_0,0),\quad m=(m_0,0,0,m_0), \quad k_0,l_0,m_0 > 0,
\end{align}
we have
\begin{equation}
 \kappa^1=\sqrt{\frac{c_4}{c_2}}k_0, \quad \kappa^2=\sqrt{\frac{c_4}{c_2}}l_0, \quad\kappa^3=\sqrt{\frac{c_4}{c_2}}m_0
\end{equation}
and
\begin{equation}
(\bm{A}_0,\bm{A}_1,\bm{A}_2,\bm{A}_3)=\frac{c_2}{c_4}(\bm{a}+\bm{b}+\bm{c},\bm{a},\bm{b},\bm{c}).
\end{equation}
The variable $\omega$ defined in Eq. (\ref{eqn:omega}) is more explicitly  given by
\begin{align}
&\omega=L\cdot x,\\
&L_\mu=\frac{\beta(\beta+2)}{2\nu^2(\beta+1)}[-\nu\beta(\beta+2)\frac{k_\mu}{\kappa^1}+\mu\beta(\beta+2)\frac{l_\mu}{\kappa^2}-2\mu\nu(\beta
+1)\frac{m_\mu}{\kappa^3}],
\end{align}
We note that $\omega$ depends on the time variable $x_0$.
                                                                                
In the case  $\mu=1$, $\nu=-1$, we have
\begin{align}
&L=(L_0,\bm{L}),\\
&L_0=2 \Lambda (\beta^2+3 \beta+1),\\
&\bm{L}=\Lambda \varpi \hat{\bm{u}},\\
&\hat{\bm{u}}=\frac{1}{\sigma} \big(\beta (\beta+2), \beta (\beta+2), 2(\beta+1) \big),\\
&\Lambda=\frac{\beta(\beta+2)}{2(\beta+1)}\sqrt{\frac{c_2}{c_4}},\\
&\varpi=\sqrt{2 \beta^2 (\beta+2)^2+4(\beta+1)^2},\\
&L^2=L^\mu L_\mu=2\beta(\beta+2)(\beta^2+6\beta+4)\Lambda^2.
\end{align}
Since  $L^2 $ for $\beta =-2-\frac{1}{\sqrt{2}}+\frac{\sqrt{5+4\sqrt{2}}}{2}$ is negative, there exists  a Lorentz transformation $ (L_0,\bm{L})\rightarrow (0,\bm{L}' )$.
Since $\omega=L\cdot x=L'\cdot x'$ does not contain the time variable $x_0'$ in
the new system, $\bm{a}$, $\bm{b}$ and $\bm{c}$ can now be regarded  as static.
For selfcontainedness, we describe the above Lorentz transformation in some detail  in the Appendix.
                                                                                
The static energy density $E$ is now given as follows. It  consists of two parts $E_1$ and $E_2$ :
\begin{align}
E = &\frac{8{c_2}^2}{c_4}[E_1 +E_2],\label{eqn:statice}\\
E_1=& 2\Big\{ ( v_1^2   \mbox{\boldmath {$a$}}^2 + v_2^2  \mbox{\boldmath {$b$}}^2 +v_3^2  \mbox{\boldmath{$c$}} ^2 )\nonumber\\
& + (2v_1 v_2 - 1)  \mbox{\boldmath {$a$}} \cdot  \mbox{\boldmath {$b$}} + (2 v_2 v_3 - 1)  \mbox{\boldmath {$b$}} \cdot  \mbox{\boldmath {$c$}} +  (2v_3 v_1 -
1)  \mbox{\boldmath {$c$}} \cdot  \mbox{\boldmath {$a$}}  \Big\} \\
E_2=&    (4v_2 v_3 - 1)  \mbox{\boldmath{$A$}} ^2 + (4v_3 v_1 - 1)  \mbox{\boldmath{$B$}} ^2+ (4v_1 v_2 - 1)  \mbox{\boldmath{$C$}} ^2  \nonumber\\
& + (2+4v_3^2-4v_2v_3-4v_3v_1)  \mbox{\boldmath{$A \cdot B$}}
 +(2+4v_1^2-4v_3v_1-4v_1v_2)  \mbox{\boldmath{$B \cdot C$}}\nonumber\\
& + (2+4v_2^2-4v_1v_2-4v_2v_3)  \mbox{\boldmath{$C \cdot A$}},
\label{eqn:ee2}
\end{align}
where  $\mbox{\boldmath{$A, B, C$}}$,$v_1,v_2$, and $v_3$  are defined by
\begin{align}
&  \mbox{\boldmath{$A$}} =  \mbox{\boldmath{$b \times c$}},\quad  \mbox{\boldmath{$B$}} =  \mbox{\boldmath{$c \times a$}},\quad  \mbox{\boldmath{$C$}} =  \mbox{\boldmath{$a \times b$}},\\
& v_1= v_2=\cosh \theta + \frac{\beta (\beta+1)}{\varpi}\sinh \theta, \quad v_3=\cosh \theta + \frac{2 (\beta+1)}{\varpi}\sinh \theta,\\
&\sinh \theta= -\sqrt{2}\frac{\beta^2+3\beta+1}{\sqrt{-\beta(\beta+2)(\beta^2+6\beta+4)}},\\
&\cosh \theta= \sqrt{-\frac{\beta^4+4 \beta^3 +6 \beta^2+4 \beta+2}{\beta(\beta+2)(\beta^2+6\beta+4)}}.
\end{align}
\subsection{Explicit solution}
To obtain a set of constants ($\mbox{\boldmath{$a$}}^2$, $\mbox{\boldmath{$b$}}^2$, $\mbox{\boldmath{$c$}}^2$) 
realizing the three conditions at the top of this section, we adopt a simplifying assumption
$\mbox{\boldmath{$a$}}^2 = \mbox{\boldmath{$b$}}^2 \equiv x$.  If we set $\mbox{\boldmath{$c$}}^2 \equiv y$, we find that $g_2$ and $g_3$ are  polynomials of $x, y$ of second and third order, respectively. 
It can be seen that many terms in $g_2$ and $g_3$ vanish when $J$ defined by
\begin{equation}
 J=\beta^4+8\beta^3+18\beta^2+16 \beta+4.
 \end{equation}
 vanishes.
We here note  that $\beta = -2-\frac{1}{\sqrt{2}}+\frac{\sqrt{5+4\sqrt{2}}}{2}$
is one of the solutions of ${J = 0}.$
Then, the condition $(g_2)^3-27 (g_3)^2 =0$ simplifies to
 \begin{align}
 &y (p_0 x+p_1y) (q_0 x^2+8q_1x y+8q_2 y^2)^2 = 0,\label{eqn:eqn30}\\
 &p_0= -74 - 52\sqrt{2}+22\sqrt{5+4\sqrt{2}}+16\sqrt{10+8\sqrt{2}},\\
 &p_1= -599 - 424\sqrt{2}+183\sqrt{5+4\sqrt{2}}+130\sqrt{10+8\sqrt{2}},\\
 &q_0= -175979 - 124436\sqrt{2}+53907\sqrt{5+4\sqrt{2}}+38118\sqrt{10+8\sqrt{2}},\\
 &q_1= -292919 - 207125 \sqrt{2}+89729\sqrt{5+4\sqrt{2}}+63448\sqrt{10+8\sqrt{2}},\\
 &q_2=-272444 - 192647 \sqrt{2}+83457\sqrt{5+4\sqrt{2}}+59013 \sqrt{10+8\sqrt{2}}.
 \end{align}
The only solution of Eq.(\ref{eqn:eqn30}) compatible with the conditions $x>0, y>0, g_2>0$ and $g_3<0$ is  
\begin{align}
&y= \tau x,\label{eqn:eqn36}\\
&\tau= \frac{C+\sqrt{D}}{F}= 0.724872\cdots,
\end{align}
where $C,D$ and $F$ are given by
\begin{align}
  &C= - 585838 - 414250\sqrt{2}+179458 \sqrt{5+4 \sqrt{2}}+126896 \sqrt{10+8 \sqrt{2}},\\
&D=726708054834 + 535418248054\sqrt{2}-303040027330 \sqrt{5+4 \sqrt{2}}\\
 & -214281658296 \sqrt{10+8 \sqrt{2}}+32820149302 \sqrt{114+80\sqrt{2}}, \\
 &F= 4\big(-272444 - 192647 \sqrt{2}+83457\sqrt{5+4 \sqrt{2}}+59013 \sqrt{10+8 \sqrt{2}} \big).
\end{align}
Utilizing the above results and together with formulae (\ref{eqn:ab}), (\ref{eqn:bc}) and (\ref{eqn:ca}), condition (\ref{eqn:cond3}) is satisfied automatically.
Now we turn to the static energy density $E$. As is seen from (\ref{eqn:ee2}), the second part $E_2$ of $E$ consists only of $\mbox{\boldmath{$A$}}$, $\mbox{\boldmath{$B$}}$ and $\mbox{\boldmath{$C$}}$.
Under the assumptions adopted here, it turns out that all of $\lim_{|\chi| \to \infty}\mbox{\boldmath{$A$}} ,\quad \lim_{|\chi| \to \infty}\mbox{\boldmath{$B$}} ,\quad \lim_{|\chi| \to \infty}\mbox{\boldmath{$C$}}$ are proportional to $q_0 x^2+8q_1x y+8q_2 y^2$. We then see that, under the choice $y= \tau x$, $E_2$ vanishes at points far apart from the plane $\chi=\mbox{\boldmath{$L$}}'\cdot \mbox{\boldmath{$x$}} = 0$. This behavior is just like that of a domain wall. On the other hand, the first part  $E_1$ of $E$ tends to a non-vanishing constant  as $\chi$ tends
to $\infty$. We show the behavior of $\mbox{\boldmath{$A$}}^2$ , $\mbox{\boldmath{$B$}}^2 = \mbox{\boldmath{$C$}}^2$ in Figs. 1 and 2. The behavior of the static energy densities  $E_1$ and $E_2$ are shown  in Figs. 3 and 4. The behavior of total static energy density $E=E_1+E_2$ is shown in Fig.5. We note that these figures are drawn under the assumption 
$x=\sqrt{y}$ in addition to (\ref{eqn:eqn36}), yielding $x=\tau, y=\tau^2$.
The baryon number density in this case is given by
\begin{align}
N_0(x) =N\Big(\frac{c_2}{c_4}\Big)^{3/2},
\end{align}
where $N$ and $n_0$ are given by
\begin{align}
&N= n_0  \frac{dK(\omega)}{d\omega},\\
&n_0 = \frac{1}{2 \pi^2}\frac{\beta^3 (\beta+2)^3}{\beta+1}.
\end{align}
The behavior of baryon number density is depicted in Fig. 6. We find that it is positive on one side of the plane and negative on the other side and the integral $\int^{\infty}_{-\infty} d\omega N_0(x)$ vanishes. We note that, for instance in the $x=\sqrt{y}$ case, we have $\int^{0}_{-\infty} d\omega N(x)=-\int^{\infty} d\omega N(x)= 0.0297476$.

We also note that another solution similar to the above one can be obtained also in the case of (2.26). In this case, we have 
$\int^{0}_{-\infty} d\omega N(x)=-\int^{\infty}_{0} d\omega N(x)= 
0.0151418$ for $x=\tau, y=\tau^2$. 

\begin{figure}[bhtp]
\centering
\includegraphics[width=9cm]{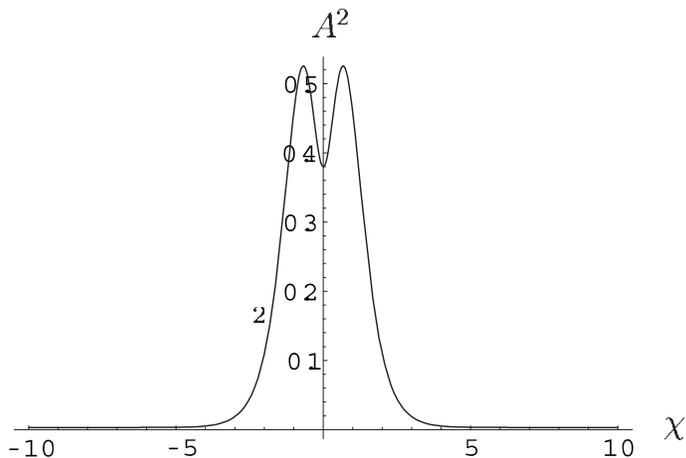}
\caption{Behavior of $\bm{A}^2$ ( $a^2=b^2=0.729476\cdots, c^2=0.532136\cdots$)}
\label{fig:behavior1}
\end{figure}

\begin{figure}[phtb]
\centering
\includegraphics[width=7cm]{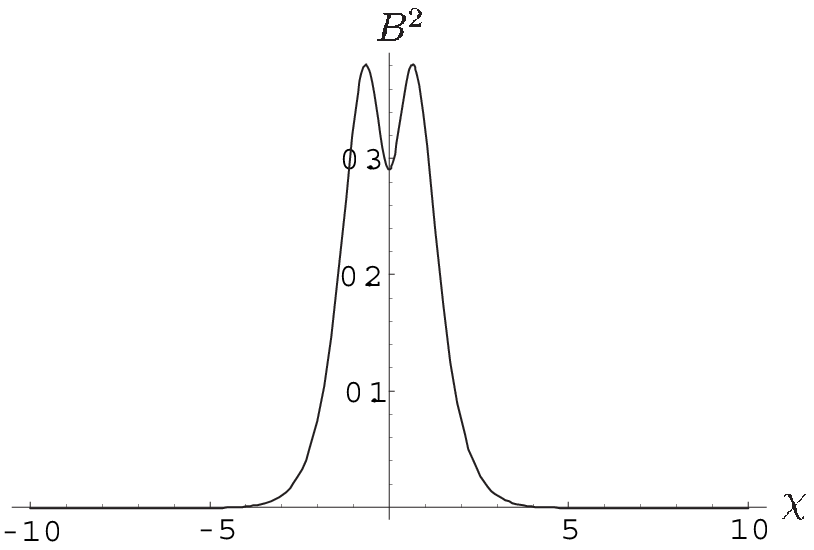}
\caption{Behavior of $\bm{B}^2 = \bm{C}^2$ ( $a^2=b^2=0.729476\cdots, c^2=0.532136\cdots$)}
\label{fig:behavior2}
\end{figure}
                                                                                                                                               
\begin{figure}[phtb]
\centering
\includegraphics[width=8cm]{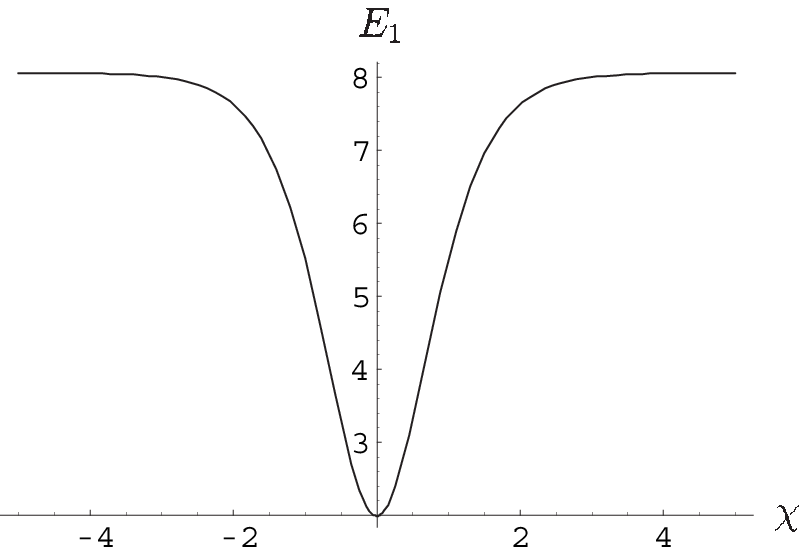}
\caption{Static energy densities $E_1$}
\label{fig:behavior3}
\end{figure}

\begin{figure}[phtb]
\centering
\includegraphics[width=8cm]{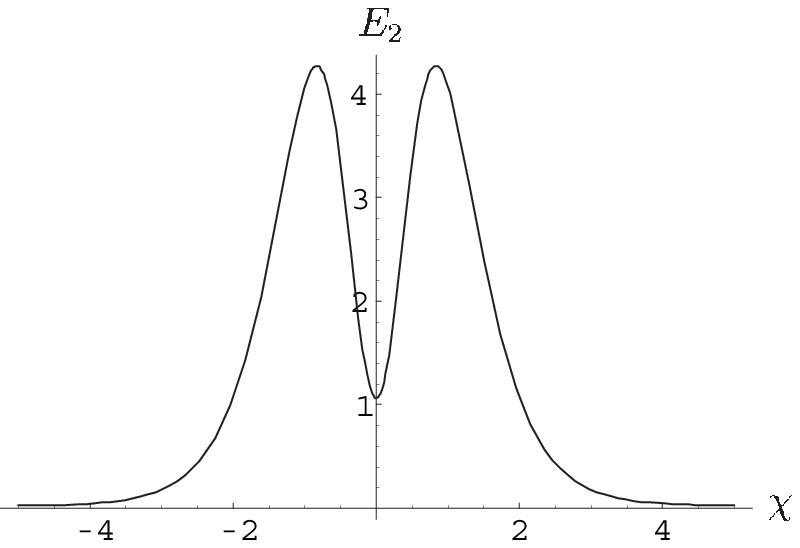}
\caption{Static energy densities $E_2$}
\label{fig:behavior4}
\end{figure}
                                                                                                                                               
\begin{figure}[phtb]
\centering
\includegraphics[width=8cm]{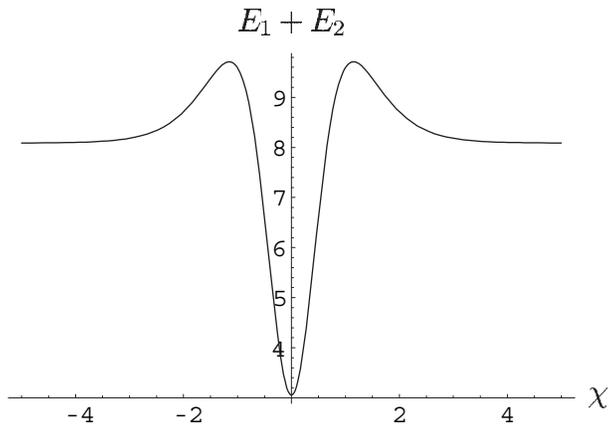}
\caption{Total static energy}
\label{fig:behavior5}
\end{figure}

\begin{figure}[phtb]
\centering
\includegraphics[width=10cm]{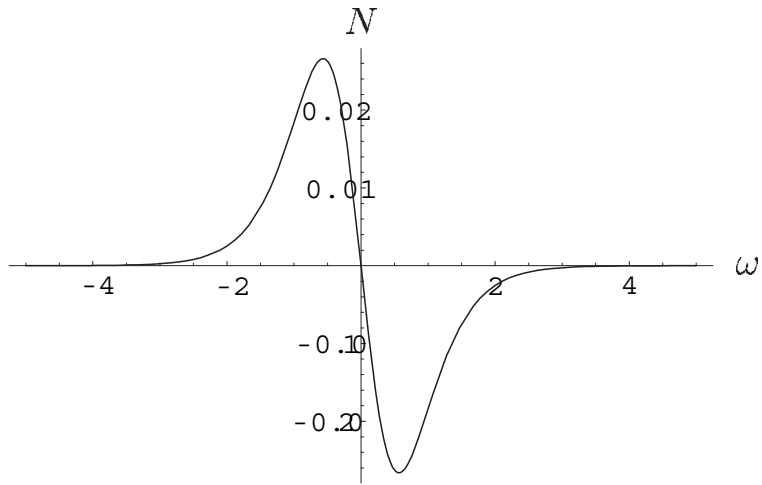}
\caption{Baryon number density}
\label{fig:behavior6}
\end{figure}
\section{Summary }\label{sec:4}
We have explored the domain wall solution of the Skyrme model. To obtain it, we have considered a limiting case of the previously obtained solution of the Skyrme model described by the Weierstrass $\wp$ function.  In the limit considered, the function $\rm{sn}(u,k)$ tends to $\rm{sn}(u,1) = \rm{tanh}(u)$ leading to a static domain wall-like solution of the Skyrme model.  Because of the complicated structure of the model, we needed to show explicitly that the above limit is indeed realizable. We have shown that there indeed exists a set of parameters which realize the limit.
The two terms constituting the static energy density of the Skyrme model were also investigated. The behavior of the term quartic
in the field variables turns  out just like that of a domain
wall : it is non-vanishing only in the neighborhood of a plane(wall) in the space. On the other hand, the term quadratic in the field variables approaches to a non-vanishing constant at points far apart from the wall. The baryon number density concentrates near the plane. It is positive in
one side of the plane and negative on the other side.  The total baryon number vanishes and the topological stability of the solution is not maintained. To obtain solutions with non-vanishing baryon number, it would be necessary to obtain solutions expressed not by $\rm{tanh}^2(u)$ but by. for example,  $\rm{tanh}(u)$.
                                                                                                                                               
\section*{Acknowledgments}
   One of the author (M. H.) is grateful to Hiroshi Kakuhata, Jun Yamashita, Shinji Hamamoto and Takeshi Kurimoto for valuable discussions.
This work is partially supported by the Science Foundation of Shanghai Municipal  Education Commission of China under Grant No.05LZ08 and the Foundation of Shanghai University of Electric Power (No.K2005-01).

\appendix
\section{Detail of Lorentz transformation}
The energy momentum tensor (\ref{eqn:energymt}) is rewritten as
\begin{align}
T_{\mu\nu}&=\displaystyle{\frac{16 {c_2}^2}{c_4}}\biggl[k_{\mu\nu}\bm{a}^2+l_{\mu\nu}\bm{b}^2+m_{\mu\nu}\bm{c}^2+r_{\mu\nu}(\bm{a}\cdot\bm{b})
+s_{\mu\nu}(\bm{b}\cdot\bm{c})\nonumber\\
&\hspace{18mm}+u_{\mu\nu}(\bm{c}\cdot\bm{a})-\eta_{\mu\nu}(\bm{a}\cdot\bm{b}+\bm{b}\cdot\bm{c}+\bm{c}\cdot\bm{a})\biggr]\nonumber\\
&+\displaystyle{\frac{8{c_2}^2}{c_4}}\biggl[4(k_{\mu\nu}\bm{B}\cdot\bm{C}+l_{\mu\nu}\bm{C}\cdot\bm{A}+m_{\mu\nu}\bm{A}\cdot\bm{B})\nonumber\\
&\hspace{18mm}+2r_{\mu\nu}(\bm{C}^2-\bm{C}\cdot\bm{A}-\bm{B}\cdot\bm{C})\nonumber\\
&\hspace{18mm}+2s_{\mu\nu}(\bm{A}^2-\bm{C}\cdot\bm{A}-\bm{A}\cdot\bm{B})\nonumber\\
&\hspace{18mm}+2u_{\mu\nu}(\bm{B}^2-\bm{B}\cdot\bm{C}-\bm{A}\cdot\bm{B})\nonumber\\
&\hspace{18mm}-\eta_{\mu\nu}(\bm{A}^2+\bm{B}^2+\bm{C}^2-2\bm{A}\cdot\bm{B}-2\bm{B}\cdot\bm{C}-2\bm{C}\cdot\bm{A})\biggr],
\label{eqn:density1}
\end{align}
where the following matrices have been made use of:
\begin{align}
&(k_{\mu\nu})=
\begin{pmatrix}
1 & 1 & 0 & 0\\
1 & 1 & 0 & 0\\
0 & 0 & 0 & 0\\
0 & 0 & 0 & 0
\end{pmatrix},
\hspace{3mm}
(l_{\mu\nu})=
\begin{pmatrix}
1 & 0 & 1 & 0\\
0 & 0 & 0 & 0\\
1 & 0 & 1 & 0\\
0 & 0 & 0 & 0
\end{pmatrix},
\hspace{3mm}
(m_{\mu\nu})=
\begin{pmatrix}
1 & 0 & 0 & 1\\
0 & 0 & 0 & 0\\
0 & 0 & 0 & 0\\
1 & 0 & 0 & 0
\end{pmatrix}\nonumber\\
&(r_{\mu\nu})=
\begin{pmatrix}
2 & 1 & 1 & 0\\
1 & 0 & 1 & 0\\
1 & 1 & 0 & 0\\
0 & 0 & 0 & 0
\end{pmatrix},
\hspace{3mm}
(s_{\mu\nu})=
\begin{pmatrix}
2 & 0 & 1 & 1\\
0 & 0 & 0 & 0\\
1 & 0 & 0 & 1\\
1 & 0 & 1 & 0
\end{pmatrix},
\hspace{3mm}
(u_{\mu\nu})=
\begin{pmatrix}
2 & 1 & 0 & 1\\
1 & 0 & 0 & 1\\
0 & 0 & 0 & 0\\
1 & 1 & 0 & 0
\end{pmatrix}.
\end{align}
We here describe some details of the Lorentz transformation from the system with $L=(L_0,\bm{L})=(L_0,-|\bm{L}|\hat{\bm{u}})$ to that with $L'=(0,\bm{L'})=(0,\varepsilon \sqrt{-L^2})$. It is   given by
\begin{align}
\begin{pmatrix}
0 &\\
-\sqrt{-L^2}
\end{pmatrix}
=
\begin{pmatrix}
\cosh\theta & \sinh\theta\\
\sinh\theta & \cosh\theta
\end{pmatrix}
\begin{pmatrix}
L_0 &\\
-|\bm{L}|
\end{pmatrix},
\end{align}
from which we obtain
\begin{align}
&\sinh \theta=\frac{l_0}{\sqrt{-L^2}}=-\sqrt{2}\frac{\beta^2+3\beta+1}{\sqrt{-\beta(\beta+2)(\beta^2+6\beta+4)}}.
\end{align}
With the help of the matrices
\begin{align}
R_1=
\begin{pmatrix}
\frac{1}{\sqrt{2}} & \frac{1}{\sqrt{2}} & 0\\
-\frac{1}{\sqrt{2}}  & \frac{1}{\sqrt{2}} & 0\\
0 & 0 & 1
\end{pmatrix}
\end{align}
and
\begin{align}
R_2=
\begin{pmatrix}
\cos\phi & 0 & \sin\phi\\
0  & 1 & 0\\
-\sin\phi & 0 & \cos\phi
\end{pmatrix},
\end{align}
we find
\begin{align}
R_2 R_1{}^t\hat{\bm{u}}=\frac{1}{\varpi}
\begin{pmatrix}
\sqrt{2}\beta(\alpha+2)\cos\phi+2(\beta+1)\sin\phi\\
0\\
-\sqrt{2}\beta(\beta+2)\sin\phi+2(\beta+1)\cos\phi
\end{pmatrix}.
\end{align}
If we set
$\cos\phi=\frac{\sqrt{2}\beta(\beta+2)}{\varpi}$,
we have
\begin{align}
R_2 R_1{}^t\hat{\bm{u}}=
\begin{pmatrix}
1 \\
0 \\
0
\end{pmatrix}.
\end{align}
So in this coordinate system, $\bm{L}$ is given by
\begin{align}
{}^t\bm{L}=
\begin{pmatrix}
\Lambda\varpi\\
0\\
0\\
\end{pmatrix}
=
\begin{pmatrix}
-|\bm{L}|\\
0\\
0
\end{pmatrix}.
\end{align}
Combining the above two transformations, we find that the static energy density is given by Eq. (\ref{eqn:statice}).
                                                                                                                                               

 \end{document}